\documentstyle[aps,prl,twocolumn,epsf]{revtex}
\begin{document}

\twocolumn[\hsize\textwidth\columnwidth\hsize\csname@twocolumnfalse\endcsname

\title{\bf Large nonzero-moment magnetic strings in antiferromagnetic
 crystals of the manganite type }
\author{E.L. Nagaev}
\address{Institute of Radio Engineering and Electronics,
 Russian Academy of Sciences\\
 Mokhovaya ul. 11, korp. 7, Moscow 101999, Russia}

\maketitle

\begin{abstract}

The magnetic strings in antiferromagnetic crystals
with the spin $S = 1 /2$ differ from
the magnetic polarons (ferrons) by the absence of the additional magnetic
moment. We show that in the $S > 1 /2$ double exchange crystals with
the antiferromagnetic $s-d$ exchange, a new type of magnetic strings
appears which possesses a magnetic moment. It is concentrated at the
center of the string, and the magnetized string is, in its essence,
the state intermediate between the string and the ferron. In 
antiferromagnetic manganites, this moment is by an order of magnitude
larger than that of a magnetic atom. Unlike the conventional ferrons,
the magnetization of the strings exists at any parameters of the
crystals under consideration. We argue that this new type of magnetic
state can be relevant to some doped antiferromagnets including
manganites.
\end{abstract}
\vskip2pc
]

\section{Introduction}

The aim of the present paper\footnote{Professor E.L. Nagaev passed
 away unexpectedly in December 2001, shortly after finishing the
 initial version of this paper. Present version was prepared for
 publication by Professor D.I. Khomskii (Groningen University).}
is to develop a theory of self-trapped charge carrier states for
 the antiferromagnetic semiconductors of the manganite type,
including manganites themselves. They are characterized by the double
exchange and by the antiferromagnetic
exchange of localized spins with the magnitude $S$
exceeding 1/2. In contrast to previous treatments, we consider here
the situation where the exchange between localized and itinerant
electrons is antiferromagnetic, or rather when they have to be
antiparallel due to the Pauli principle.

As is well known, a background
antiferromagnetic ordering does not ensure the minimum energy of
charge carriers. That is why the charge carriers tend to destroy the
antiferromagnetic ordering and replace it by another ordering or by a
disordered state, at least in their vicinity. The destruction can be
either static or dynamic (we neglect the translational motion of ferrons
or magnetic strings discussed below, as their bandwidths are very
small and the Anderson localization should take place in real crystals).

The tendency toward the static local
destruction of an antiferromagnetic order manifests itself in the
formation of the magnetic polarons (ferrons) when a free charge
carrier becomes self-trapped by a ferromagnetic microregion
permanently existing inside an antiferromagnetic host \cite{1,2}.
 Hence their main feature is  an excess magnetic moment as compared
 with the moment of the electron generating the ferron state.

In principle, the ferrons are possible for the systems without the
double exchange independently of the sign of the $s-d$ exchange
integral, and in the double exchange systems with the ferromagnetic
$s-d$ exchange. (To shorten notation, we speak below about $s$- and
$d$-electrons, meaning by $s$ the mobile electrons, and by $d$ -
the localized ones. In reality  the mobile (``$s$'') elecrons may
also originate from the same $d$-shell. Thus, for example, in
manganites they are the $e_g$-electrons or holes, whereas our
``$d$''-electrons are the $t_{2g}$ ones). The conditions for the
existence of free ferrons are rather stringent even at $T = 0$. For
example, the N\'eel point $T_N$ should be not very high. But there
are also bound ferrons which correspond to an electron trapped by the
donor impurities. As discussed e.g. in \cite{2,3}, for them the excess
magnetization is nonzero for any N\'eel point though it decreases
with increasing $T_N$. If a region with the complete ferromagnetic
ordering  is impossible, then a canted antiferromagnetic ordering
arises close to the impurity \cite{3,4} (see also references to papers
of other authors in review article \cite{3}).

 The tendency toward the dynamic destruction of antiferromagnetic
order manifests itself in the formation of magnetic strings (in the
pioneering paper \cite{5} they were called quasioscillators).
Strings are traces of wrong spins left on the trajectory of an
extra electron or hole when it moves in an antiferromagnetic
background in a situation described by the nondegenerate Hubbard
model. In this state, a charge carrier moves by interchanging with the
neighboring spins, so that as a result the net magnetization does
not change, remaining essentially zero. (The formation of such
strings and the resulting electron or hole confinement was
rediscovered \cite{6,7} much later than the first paper \cite{5}
appeared, after the interest to such model was raised by the
discovery of high-temperature superconductivity).

For this quasiparticle, oscillations of the magnetic disorder
are accompanied by the oscillations of the electron density.
 This disordering excludes the possibility of the free charge
 carrier motion, and hence
the standard scenario of the ferron formation is impossible here.
The existing theory of the strings \cite{5,6,7} is related to the
systems described by the Hubbard Hamiltonian that are equivalent to
 double-exchange systems
 with the magnetic atom of spin $S = 1 /2$ and the negative
(antiferromagnetic) sign of the $s-d$ exchange integral, which ensures
the zero total spin of an atom occupied by an $s$-electron \cite{2}.
As explained above, unlike ferron, the magnetic string with
$S = 1 /2$ does not possess an additional magnetic moment as compared with
the moment of the self-trapped electron.

Meanwhile, the case of spins $S$ exceeding 1/2 is important for real
materials, but was not investigated yet. It can be described by
the $s-d$ model with the antiferromagnetic on-site $s-d$ exchange.
The free motion of the charge carriers
is again impossible in this case, and the strings arise, see Fig.1
below. However, as we will show, the strings in such systems differ
from the strings in a $S = 1 /2$ magnetic system since they can
have an excess magnetic moment. This moment will be concentrated in a
region around the equilibrium point, i.e. the core. The central part
of this object can be ferromagnetic (ferromagnetic polaron, or
ferron), whereas the motion of charge carriers outside this core
region occurs in a string-like fashion. Thus, the resulting state may
be visualized as a combination of the inner ferron \cite{1} and outer
string \cite{5} regions.

As will be shown below, the string excess moment exists for
arbitrary parameters
of the system under consideration. In the crystals with sufficiently
 low N\'eel temperatures $T_N$, a microregion with the complete
 ferromagnetic ordering should exist. Hence such a state
 may be interpreted as a mixed string-ferron state.
 But for the crystals with sufficiently high $T_N$, the complete
 ferromagnetic ordering should
be replaced by the canted antiferromagnetic one (the canted ordering arises
since the antiferromagnetic superexchange in the high-$T_N$ 
systems is too strong to be completely suppressed by the
ferromagnetic indirect exchange via the bound charge carrier).
The existence of  the excess magnetic moment at any $T_N$
 stems from the fact that the string state is a bound state
resembling the state of an electron trapped by a donor. The fact that
the force in the former case is
quasielastic and not the Coulomb is nonessential, as both these forces
cause the electron confinement. In other words, the electron
localization for the magnetized string  is caused not by the
arising magnetized region but by the quasielastic force.

 The calculations carried out below show, in particular, that such
quasiparticles should exist in manganites \cite{3,8,9} when they
are antiferromagnetic, and their magnetic moment should be an order
of magnitude larger than the moment of a separate Mn$^{3+}$ ion.
(Strictly speaking, our treatment will be
carried out for the slightly doped simple two-sublattice
antiferromagnet with the $G$-type structure, whereas undoped
LaMnO$_3$ has the layered $A$-type structure. But the general
physical mechanism, considered in this paper, should work in this
case as well, albeit with some modifications; besides, there exist
many other magnetic systems with the $G$-type structure which can be
doped and which could be described by our model.)

The mobility of the magnetized strings, like that of the
conventional ferrons, is too low to affect the conductivity. However,
they can produce a significant effect on magnetic properties
 of the system.  As is well known, the existence of the "ready
at hand" large magnetic moments in an antiferromagnetic semiconductor
 leads to a steep rise in the magnetization under
a weak external magnetic field. This effect is often observed
 experimentally. Certainly, in a nondegenerate semiconductor
 all these quasiparticles  are bound to the
donor (acceptor) impurities at $T \to 0$. But, at finite temperatures, the
ferrons can become free, i.e. become  located far from impurities. Just these
ferrons contribute to the jump in the magnetization. For this reason, the
evaluation of their moments is important for the interpretation of the
experimental data.

\section{Model}

The problem of the energy spectrum in the double exchange
systems is treated using  the $s-d$-model  as a basis:
$$H_{sd} = -t \sum a^{*}_{{\bf g} \sigma} a_{{\bf g+\Delta}\sigma}
 -A \sum \left({\bf sS_{g}}\right)_{\sigma \sigma'}
 a^{*}_{{\bf g} \sigma} a_{{\bf g}\sigma'}$$

$$- \frac{I}{2} \sum {\bf S_{g}S_{g+\Delta}}, \qquad (1)$$
where  $a^{*}_{{\bf g} \sigma},  a_{{\bf g}\sigma}$
   are the $s$-electron operators corresponding to the conduction
electrons or holes at atom {\bf g} with spin projection
 $\sigma$,{\bf s} is the $s$-electron spin operator, ${\bf S_{g}}$ 
 that of the $d$-spin
of atom {\bf g}, ${\bf \Delta}$ is the vector connecting
 the first nearest neighbors.  The crystal structure is assumed
 to be simple cubic. As usually, the
inequality $t  \gg  |I|S^2$
must be met as the hopping integral $t$ is of the first order of
magnitude, and  the $d-d$ exchange integral $I$ of the second order in
the small $d$- orbit overlapping ($I$
should be negative to ensure the antiferromagnetic ordering).

In what follows, the condition of the double exchange  $12t \ll |A|S$
will be assumed to be met, with the $s-d$ exchange integral $A$ being
negative. Under these conditions, in the zero approximation in
$W/AS$  $(W = 12 t)$ , the charge carrier is fixed at one of the magnetic
atoms, their total spin being $S - 1/2$.

It is necessary to establish first when model (1) with $A < 0$ is
realistic. To begin with, in reality, the double exchange takes place when a
charge carrier is an extra or deficient electron in the $d$-shell of 
transition metal compounds. According to the Hund rule, the $d-d$ exchange is
ferromagnetic; the same should be true for the exchange between the conduction
electron and the rest of the $d$-spins, which form the localized spins of the
magnetic atoms. According to the Pauli principle, the total spin of the
atom, occupied by a conduction electron at a certain moment of time,
 cannot exceed 5/2. Hence the charge carrier spin is parallel to
the total spin of the other atoms only for $S \le 2$.

If $S \ge 5/2$, despite the ferromagnetic on-site exchange, the spin of
the charge carrier is antiparallel to the total spin of the rest of the 
$d$-electrons. Hence in this case an antiferromagnetic $s-d$ exchange integral
should be chosen in the $s-d$ model. The same should be done, if the
 charge carrier is a hole, and  $S \le 5/2$. Really, if the $s-d$ Hamiltonian (1) is written  in terms of the
hole operators, the energy of the hole is opposite in sign to the
corresponding electron energy. Thus, the assumption of an negative
 $A$ is justified for $S$ not less than 5/2 or not larger than 5/2
 if the charge carriers are the electrons or the holes, respectively.
 In particular,  the latter condition is
met for  the manganites with the hole conductivity and  $S = 2$ for the
Mn$^{3+}$ ions.

It should be noted that in the case of the electrons the
negative $A$ looses the meaning of the Hund's exchange integral and reflects
only the Pauli principle. But this is nonessential, as quantity
$A$ enters only in the zeroth approximation in $W/AS$ via an additive
constant $A(S+1)/2$, which can be omitted. The point is that the
quantity $A$ does not enter the terms of the first approximation in
the effective Hamiltonian (2) which describe the double exchange.

This consideration can be readily generalized to the case
when the crystal field splits the $d$-level so strongly that the gap
between the sublevels exceeds the $s-d$ exchange energy, and all
 the $d$-electrons occupy states inside the lowest sublevel.
 Then the value of the maximum spin of the entire $d$-shell should
 be replaced by that of the lowest $d$-sublevel when
considering the problem in a way similar to that just used. If the lowest
sublevel is completely filled, the role of the entire $d$-shell is played
by the upper level.

 As mentioned in the Introduction, to simplify the treatment, the
magnetic ordering is assumed to be staggered instead of the layered
one which is realized in the undoped manganites. Conventional strings
in layered antiferromagnetic structures with $S =1/2$ were
considered in Ref.~\cite{9}. It should be pointed out that for $A>0$
the charge carrier motion is band-like though at the
antiferromagnetic ordering the band is considerably narrower than for
the  ferromagnetic that. Conventional ferron states are possible in
this case~\cite{1,2}.

\section{Nonmagnetized magnetic strings}

Before carrying out real calculations, we first explain qualitatively
the origin and the nature of the string states in our case. Consider
an antiferromagnet with the localized spins $S$ forming a simple
two-sublattice N\'eel structure, and let us put an extra electron at
site $a$ with the spin $S^z = S$. In the situation considered (e.g.
$S=5/2$), it would necessarily have spin down --- or, more accurately,
the total spin of this state will be $S-1/2$. In our model,
only this extra electron (``s''-electron) can move to a neighboring
site. But if we consider the resulting situation classically, this
extra electron with spin down will not be able to move at all,
because at the neighboring sites there already exists spins down
(e.g. $S^z = 5/2$), so that the Pauli principle forbids such motion.

\begin{figure}[tbp]
\epsfxsize= .75\hsize  \vskip 1.0\baselineskip
\centerline{\epsffile{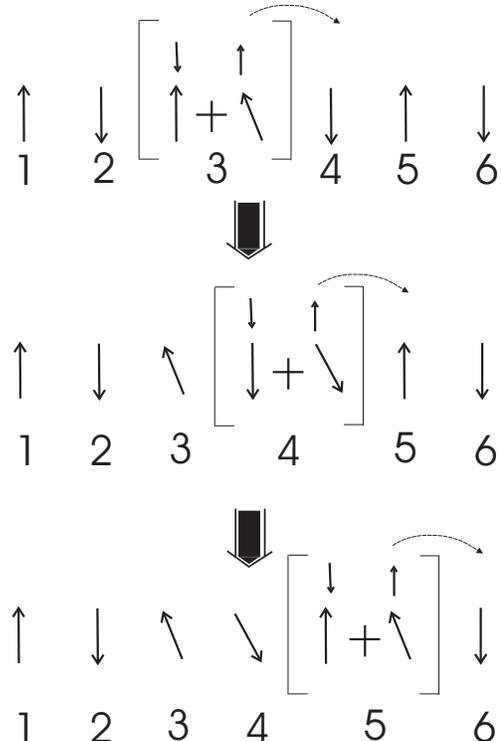}}
\caption{Schematic picture explaining the formation of a string
(deviations of the localized spins (large arrows)) in the process of
motion of the conduction electron (small arrow). The exchange
interaction of these tilted spins with the surrounding
antiferromagnetic spins in two- and three-dimensional cases
 (not shown) will lead to an energy increase proportional
 to the length of the electron trajectory.}
\label{Fig.1}
\end{figure}
 
However the situation changes if one takes into account the quantum nature of spins.
The state with the total spin $S_{tot}=S-1/2$ at an initial site $a$
is not necessarily that with the localized spin $S^z = S$ and the
spin of the s-electron $s^z = -1/2$: it also has another component,
with the localized spin $S^z = S-1$, and $s^z = +1/2$,

$$ | S_{tot} = (S-1), S_{tot}^z = (S-1) \rangle$$
$$ = \alpha \; | S^z = S, s^z = -1/2 \rangle
 + \beta \; | S^z = (S-1), s^z = 1/2 \rangle $$

The $s$-electron can hop to a neighboring site $b$ owing to the second
part of this wave function. But then there will be a spin
deviation left at the initial site $a$: instead of the state with
$S^z = S$, as in the beginning, now it will have $S^z = S-1$.
This process can be repeated, and as a result the charge
carrier will move, but leaving the trace of spin deviations left at
its trajectory. This is illustrated in Fig.1.

As we see, this process is quite similar to the one, which leads
to a string formation in a nondegenerate Hubbard model \cite{5}, see also
\cite{10}. The difference is that in this case, it is not a completely
reversed spin which will be left on a trajectory of a charge carrier,
but the spin deviation of a large spin $S$. But, similar to the
case of \cite{5}, one immediately sees that the trace of wrong spins would
cost us an energy of the exchange interaction  with the neighbors
(in two- and three-dimensional systems) proportional to the length of the
trajectory, which would finally lead to a confinement of the electron
(if one ignores the spin-flip processes and the Trugman
trajectories \cite{11}, which give only a small contribution in the case
of large spin $S$).
However, in addition to that, we will show that the string formation
in this case can be accompanied also by the formation of the
magnetized ``core'', so that the resulting object would have a
nonzero magnetic moment.

We go now to the mathematical treatment of this problem. First,
 we consider the case when the magnetic moment of the string
 is negligibly small. We use the special version of the
perturbation theory in $W/AS$ developed in Refs.~\cite{2,12}.
It makes possible to construct an effective electron-magnetic
Hamiltonian $H_{ef}$ for the system under consideration, in which
the $d$-spins are considered as quantum operators. This
Hamiltonian
is exact to the first order in  $AS/W$, and no additional
approximations (mean field etc.) are used.

To construct the Hamiltonian, an effective
spin of magnitude $S$ is ascribed to each magnetic atom,
independently of its being empty or occupied by an $s$-electron.
 In the lastcase, it formally increases the number of spin degrees
 of freedom by 1 as compared with their initial real number.
 But the structure of the effective
Hamiltonian ensures vanishing of the contribution from the extra
degree of freedom. As for the charge carriers, they are treated as
spinless fermions with operators
 $ c ^{*}_{{\bf g}}, c_{{\bf g }} $. This Hamiltonian has the form

$$H_{ef} = \frac{A(S+1)}{2} \sum c ^{*}_{{\bf g}} c_{{\bf g }} -
t \sum F({\bf S_{g}, S_{g+\Delta}}) c ^{*}_{{\bf g}} c_{{\bf
g+\Delta }} -$$
$$- \frac{I}{2} \sum {\bf S_{g}S_{g+\Delta}}, \qquad (2)$$
$$ {\rm with} \qquad (2S +1) F({\bf S_{g}, S_{g+\Delta}}) = $$
$$\sqrt{(S +
S^{z}_{\bf g})
 (S + S^{z}_{\bf g +\Delta})}
+ \frac{1}{\sqrt{ S  + S^{z}_{\bf g} }}  S^{+}_{\bf g} S^{-}_{\bf g
+\Delta}
\frac{1}{\sqrt{ S  + S^{z}_{\bf g + \Delta} }}$$
In Eq.(2) small terms of the order of $IS^{2}/t$ describing a change in the
$d-d$ exchange of the atom bearing the charge carrier are omitted. An
explicit expression for them is presented in Refs. \cite{12}.

In what follows, the upper and
lower  bounds for the nonmagnetized string energy will be found
 which allows to
make the string theory more general and more accurate as compared
with previous theory for  $S$ = 1/2.  Calculations show that two
quite different approaches lead to results which are very close.
Hence both these approaches are reliable at the actual values
 of parameters.

We begin with the carrier located at the central atom ${\bf 0}$ 
 when the ordering is complete antiferromagnetic.
This means that the spin projection of each atom is maximum in the
 local coordinate system, in which the $z$-axis coincides with
the moment of the sublattice of this atom.
As is seen from Eq. (2), only such hole transitions to the neighboring
atoms are possible, when at the atom left by the hole, the deviated
spin remains with the  spin projection  $S -1$ in the local coordinate
system. At the next step the second deviated spin arises.
 and so on. As a result, the
motion of the hole along a path beginning with the central atom
is accompanied by the deviation
of spins from the moments of their sublattices along this path,
see Fig.~1.

The wave function of the system is sought in the form, in which the
possibility of the $d$-spin   deviations more than by 1 is ignored as their
relative weight is small  ($ \propto 1/z$ where $z$ is the coordination
number):
$$\Psi_s = \sum _{i=0, {\bf \Delta}_1...{\bf \Delta}_i} X ({\bf \Delta}_0,
{\bf  \Delta}_1,
.{\bf \Delta} _i)  c^{*}_{{\bf \Delta}_0 +{ \Delta}_1 + ...{\bf
\Delta}_{i}}|0>\times$$
$$ \prod_{k
=0}^{i-1}  \frac{S^{-}_{{\bf \Delta}_{0}+ {\bf \Delta}_{1} + ...{\bf
\Delta}_{k}}}
{\sqrt{2S}} \prod_{\bf f} \delta(S^z_{\bf f},S), \qquad  {\bf \Delta}_0  \equiv
{\bf 0}
\qquad  (3)$$
written using the relationship (see, e.g., Ref. \cite{2})
$$S^{-}\delta(S^z,m)= \sqrt{(S+m)(S - m + 1)}\delta(S^z,m-1).$$
Here $|0>$ is the spinless fermion vacuum wave function,
 $\delta(S^z,m)$ is the $d$-spin wave function with the spin projection $m$
equivalent to the Kronecker symbol $\delta_{S^z,m}$.
To simplify the form of  $\Psi_s$ in Eq. (3),  unlike  the Hamiltonian
(2),  the local coordinate system is used
for each atom, in which
the $z$ axis coincides with  the  moment of the sublattice to which the
spin belongs. When one calculates  the matrix elements of Hamiltonian (2)
with the use of $\Psi$ (3), one should use the common coordinate
system, for example, corresponding  to the sublattice I, and then for
the atoms of the sublattice II  one should change the sign of the  $d$-
spin projection and  replace $S^{-}$ by $S^{+}$.

The string wave equation is obtained from Eqs. (2), (3) in the
form 
$$(E -nL) X({\bf \Delta}_0,{\bf \Delta}_1, ...{\bf \Delta}_{n}) =$$
$$ = - t_s \sum_{{\bf
\Delta}_{n+1}} X({\bf \Delta}_0, {\bf  \Delta}_1
..{\bf \Delta}_n, {\bf \Delta}_{n+1}) ,\qquad (4)$$
$$t_s = \frac {t \sqrt {2S}}{2S + 1}, \qquad L = |I|S(z-2)$$
where $L$ is the change in the $d-d$ exchange energy
due to the spin deviation. Strictly speaking, this expression
 is valid for $n > 1$ if the paths are not self-crossing and self-
contacting
but the number of such paths is  very small (just they lead to the
motion of the string center over the crystal \cite{13}). The same equation
can be obtained by a direct expansion of the wave function
 $\Psi_s$ in the eigenfunctions of the second term of the Hamiltonian (1).

To solve Eq. (4), an effective wave function is introduced for the
string by summing over the paths, which lead to the same point:

$$\phi({\bf f}) = \sum_n \phi_n({\bf f}), \qquad$$
$$ \phi _n({\bf f}) =
\sum _{{\bf \Delta}_1...{\bf \Delta}_{n}} X({\bf \Delta}_0,{\bf  \Delta}_1,
..{\bf \Delta}_n)
\delta({\bf  \Delta}_1 + ...{\bf \Delta}_n,{\bf f}), \qquad (5)$$
where $\delta({\bf x,y})$ is tantamount to the three-dimensional
Kronecker symbol.

Multiplying both sides of Eq. (4) by $\delta({\bf  \Delta}_1 + ...{\bf \Delta}
_n,  {\bf f})$ and summing over all ${\bf \Delta_i}$ and $n$, one obtains:
$$\left[ L \sum_{n}n \phi_n({\bf f}) - E_s \phi({\bf f})\right] = t_s
\sum_{{\bf \Delta}} \phi({\bf f+ \Delta} ). \qquad (6)$$

To obtain the lower bound for the string energy, one should
assume that the hole reaches the atom {\bf f} using the shortest way,
i.e. one sets
$$\sum_{n}n \phi_n({\bf f}) = f  \phi({\bf f} ) \qquad (7)$$
(for the one-dimensional case this equation is exact).

The effective mass approximation is used. For the spherically
symmetric potentials the  ground state wave
function is given by \cite{14}
$$\phi(r) = \frac{\chi(r)}{r}, \qquad  \chi (0)= 0\qquad (8)$$

As a consequence of Eqs (7), (8), one arrives at the wave
equation for the wave function $\chi_l$ corresponding
to the lower bound:
$$ \left (-\frac{1}{2m_s}\frac{d^2}{dr^2} + \frac{L}{a}r - 6t_s -
E_{s}^{l}\right)\chi_l(r) = 0, \qquad (9)$$
where  $\hbar = 1$, $a$ is the lattice constant, $z$ = 6 is the
 coordination number,
 $m_s = 1/(2t_s a^2)$
is the effective mass of the hole bound to the fixed center of the string
state and has nothing in common with  the motion of the quasiparticle
over the crystal.

Eq. (9) can be solved exactly, cf.~\cite{5}. The ground state is
described by the wave function
$$\phi_l(r) = \frac{Ai[r/\rho -
a(E_{s}^{l}+6t_s)/\rho L]}{r}, \qquad$$
$$ \rho = (a/2m_sL)^{1/3} = a
(t_s/L)^{1/3}, \qquad (10)$$
where $Ai(r)$ is the Airy function, $\rho$ is 
the string radius. According to Eq. (8) 
the energy should be found
using the maximum negative value of zeroes of the Airy function
presented in Ref.~\cite{15}. Then  one obtains the lower bound for
the string energy:
 $$E_{s}^{l} = -6t_s + 2.3 L^{2/3}t_s^{1/3}. \qquad (11)$$
 This result is again similar to a nondegenerate case ($S=1/2$)
\cite{5}. For the typical values of the manganite parameters $t_s$
about 0.1 eV (the initial band width $2zt_f$ = 1.2 eV, $S$ = 2) and
$L$ about 0.01 eV the string radius $\rho$ amounts to about 2
lattice constants, and $E_{s}^{l}$ = - 0.35 eV (this energy is
counted from the band center). As is seen from this estimate, the
lower bound for the string energy is rather close to the band bottom
of a particle with the bandwidth $2zt_s$.

To obtain the upper bound for the string energy, it is sufficient to
assume that the hole reaches the atom {\bf f} by random walks. Then
for $n$ sufficiently  large the probability to reach the site ${\bf f}$
after $n$ steps  is \cite{16}
$$p_n({\bf f}) = \frac{1}{(2 \pi n)^{3/2}}{\rm exp}\left (-
\frac{f^2}{2n}\right). \qquad (12)$$
One obtains the following values for the averaged values of $f$ and
$f^2$ from Eq. (12):
$<f> = 1.596 \sqrt {n},  <f^2> = 3n$. Hence, the dispersion is
sufficiently small, and  to find the upper bound for the string energy
$E_{s}^{u}$, one should solve the following wave equation:
$$ \left (-\frac{1}{2m_s}\frac{d^2}{dr^2} + \frac{L}{2.5 a^2}r^2 -
6t_s - E_{s}^{u}\right)\chi_u(r) = 0. \qquad (13)$$
 The solutions of
Eq. (13) with the boundary condition  (8) are the wave functions of an
one-dimensional oscillator with uneven quantum numbers. Hence the
ground-state value of this number is 1, and the zero-point energy of
a three-dimensional oscillator
 is not $\omega/2$ but $3\omega/2$. This yields:
$$E_{s}^{u}= -6t_s +  1.9(t_sL)^{1/2}. \qquad (14)$$
For the parameters used above, $E_{s}^{u} = - 5.2 t_s$.
The fact that  $E_{s}^{u}$ and  $E_{s}^{l}$ presented
above virtually coincide is a consequence of the fact that, strictly
speaking,  Eq. (13) is valid for very large $f$, i.e., for small $L/t_s$, and
its value adopted here is not very small. If one takes the $L$ value
an order of magnitude less, then $E_s^u$ exceeds $E_s^l$ by 0.1 $t_s$. In
any case, the approach based on (7) ensures a reasonable accuracy at actual
parameter values. The reason for it is the fact that longer paths
to the same point correspond to larger energies of the magnetic disordering,
and hence their weight is small as compared with the shortest
path.  The approach based on (7) will be used for
investigation of the magnetized strings in the next Section.

\section{Magnetized magnetic strings}

Keeping in mind that according to Eq. (4) $t_s \propto (2S)^{-1/2}$, one
sees that the string state disappears in the classical limit $S \to \infty$.
Hence, in this case, the only way
to allow at least for some motion of an electron to decrease its
kinetic energy, would be by creating the ferron state.
But we can combine both these possibilities by making a ferromagnetic
microregion in the centre of the string. Consequently, we should
treat two problems: the motion of an electron inside a ferromagnetic
microregion of radius $R$ (to be determined by the minimization of
the total energy), and the string-like motion outside this region. The
solutions should be matched at the boundary of the ferron.

As is seen
from Eq. (2), the effective hopping integral for the  ferromagnetic
ordering is 
$$t_f = \frac{2St}{2S + 1},$$
i.e., exceeds $t_s$ by a factor of $\sqrt{2S}$. Hence,
 the ferromagnetic region is a potential
well for the carrier of a depth $$U = z (t_f - t_s) = z \frac{2St}{2S
+ 1}\left(1 - \frac{1}{\sqrt{2S}} \right)\qquad. (15)$$

As follows from Eq. (15), the string magnetization is possible
 only for $S > 1/2$.

The state of the charge carrier inside the spherical potential well of a
radius $R$ with a  ferromagnetic ordering is found from the equation
$$ - t_f (z+ a^2 \Delta)\psi  = E_{ms}\psi  \qquad (r < R) \qquad (16)$$
where $E_{ms}$ is the magnetized string energy.
As for the antiferromagnetic region outside the sphere of the radius
$R$, the same approximation as in Eq (9) will be made:
 $$ [- t_s (z+a^2\Delta)+ \frac{L}{ a}( r - R) ]\psi = E_{ms} \psi,
 \qquad (r > R). \qquad (17)$$
The ground-state solutions of Eqs (16) and (17) are,
respectively (Eq (8) is used): $$\chi  = {\rm sin} pr, \qquad  tp^2
a^2=  (E_{ms} + zt_f), \qquad (r <R) \qquad (18)$$

$$\chi = Ai(-\xi),$$
$$ \xi = \left( -\frac{r - R}{a} -\frac{U}{L}+
\frac{p^2 a^2 t_f}{L}\right)\left(\frac{L}{t_s}\right)^{1/3}.
\qquad (19)$$

An equation (boundary condition) describing the matching of the wave
functions outside and inside the  ferromagnetic region has the form

$$ap\,{\rm cot}(pR) = \frac{L^{1/3}}{ t_{s}^{1/3}}\frac{dAi(-\xi
)/dr|_{r=R}} {Ai(-\xi)|_{r=R}}. \qquad (20)$$

To calculate the right-hand side of Eq. (20), the asymptotic expression
for $Ai(-\xi )$ will be used, which is valid at $\xi \to - \infty$:
$$Ai(-\xi) = \frac{1}{2 |\xi| ^{1/4}}{\rm exp}\left( -\frac{2 |\xi|
^{3/2}}{3}\right). \qquad (21)$$
Formally this expression is valid for $\xi \gg 1$ but practically
is sufficiently accurate beginning with $\xi = 1$  \cite{15}.
Then one can rewrite Eq. (20) in the form
$$ap\, {\rm cot}(pR) = -  (2S)^{1/4}(V - a^2p^2)^{1/2}, \qquad V =
\frac{U}{t_f}. \qquad (22)$$
 Eq.\ (22)  makes it possible to find $p$ as a function of $R$ and
then to find  the $R$ value which gives the  minimum of the total
energy of the system:
$$E_{ms} = - 6t_f + p^2 a^2 t_f + \frac{4 \pi
R^3 D}{3 a^3}, \qquad $$
$$D =  z|I|S^2 =\frac{3LS}{2},  \qquad (23)$$
where $D$ is the $d-d$ exchange energy spent for formation of the
ferromagnetic microregion (per atom). Technically, it is more
convenient to fix a $p$ value and to find the lowest $R$ value for
which the Eq.~(22) is met.

Numerical calculations were carried out for $S = 2$.
 In the range of  $R/a$ values between 4.81 and 1.08,
 the solutions of Eq. (22)  can
be presented approximately in the form
$$ap(R) = \frac{3.4}{0.9 + R/a}. \qquad (24)$$
As should be the case, keeping in mind that 3.4 is sufficiently close to $\pi$,
one arrives at the same expression for $R \gg a$ as for the ferron
energy found in Refs.\ \cite{1,2}. But quite different is the case of
$R$ comparable to $a$. Then the condition for the magnetized string
existence differs from that for the ferron existence drastically.

The minimization of the energy (23), (24) with respect to $R$ yields an
expression for the energy of the $d-d$ exchange $D$ as a function of $x = R/a$:

$$\frac{D(x)}{t_f} = \frac{1.84}{(0.9 + x)^3x^2}. \qquad (25)$$

It shows
that a completely  ferromagnetic region arises when  $D/t_f < 0.27$. Then $R
=a$, and the magnetized string energy (23) is lower than the
string energy (9) though the difference is small. Energy $E_{ms}$ decreases
 and the size $R$ of the ferromagnetic region increases
 with diminishing ratio $D/t_f$.
For example, for $D/t_f$ = 0.01 energy  $E_{ms}$ is - 3.45 $t_f$
= - 6.9$t_s$ and
the radius $R$ is about 2$a$. This magnetized string energy is lower than
the lower bound for the nonmagnetized string energy (9) equal to -5.83 $t_s$.

Numerical estimates for the manganites can be obtained using the well-known
mean field expression for the N\'eel point $T_N = D(S+1)/3S$, its experimental
value 140 K and  $t_f$ is about 0.1 eV. Then one obtains $R$ close to $a$.
This means that a free string-ferron has a magnetic moment close to $zS$.
In other words, it should be an order of magnitude larger than the moment of a
Mn$^{3+}$ ion.
If ratio $D/t_f$ is larger than 0.27, formally $R$ is less
than $a$; this means that the magnetized region is rather a canted
antiferromagnetic, and not a ferromagnetic one. The situation
resembles that for the bound ferrons \cite{3,4}. But it seems not very
actual.

\section{Conclusions}

Summarizing, we considered in this paper the motion of charge carriers
in doped antiferromagnets in case when the localized electrons have
total spin $S > 1/2$ and, due to the Pauli principle, spins of
doped electrons or holes are opposite to the localized spins (i.e.\
when the total spin on a site with an extra electron or hole is
$S-1/2$). We have shown that the motion of charge carriers in this
case is possible by forming strings --- traces of spin deviations.
Such string states by themselves would not have any extra
magnetization. However the better state, with lower energy, may be
created in this case: the central part of this state will be a
ferromagnetic polaron (ferron) or a strongly canted region, and the
motion of an electron outside this region proceeds in a string-like
fashion. Thus, this state combines the features of both the ferron
states \cite{1} and of the string states \cite{5}. The extra magnetic
moment, associated with each charge carrier in this case, could
significantly modify magnetic properties of such systems. We believe
that the physics described above may be relevant for the low-doped
manganites; there may be also other materials belonging to this
class.

\section*{Acknowledgments}

This work was supported in part by Grant No.\ 01-02-97010 of
the Moscow Region Government --- Russian Foundation for Basic
Research, by Grant INTAS-97-open-30253, and by the agreement with
the Russian Ministry of Science and Industry.


\end{document}